\documentclass[aps,pre,twocolumn,showpacs,groupedaddress]{revtex4}
\usepackage[dvips]{graphicx}

\begin{document}

\title{Unified nonequilibrium dynamical theory for exchange bias and training effects}

\author{Kai-Cheng Zhang and Bang-Gui Liu}
\address{Institute of Physics, Chinese Academy of Sciences,
Beijing 100190, China} \address{Beijing National Laboratory for
Condensed Matter Physics, Beijing 100190, China}

\date{\today}

\begin{abstract}
We investigate the exchange bias and training effects in the FM/AF
heterostructures using a unified Monte Carlo dynamical approach.
This real dynamical method has been proved reliable and effective
in simulating dynamical magnetization of nanoscale magnetic
systems. The magnetization of the uncompensated AF layer is still
open after the first field cycling is finished. Our simulated
results show obvious shift of hysteresis loops (exchange bias) and
cycling dependence of exchange bias (training effect) when the
temperature is below 45 K. The exchange bias fields decrease with
decreasing the cooling rate or increasing the temperature and the
number of the field cycling. With the simulations, we show the
exchange bias can be manipulated by controlling the cooling rate,
the distributive width of the anisotropy energy, or the magnetic
coupling constants. Essentially, these two effects can be
explained on the basis of the microscopical coexistence of both
reversible and irreversible moment reversals of the AF domains.
Our simulated results are useful to really understand the
magnetization dynamics of such magnetic heterostructures. This
unified nonequilibrium dynamical method should be applicable to
other exchange bias systems.
\end{abstract}

\pacs{75.75.+a.75.20.-g,75.60.-d,05.70.Ln}


\maketitle

\section{ Introduction}

Usually, when the heterostructure consisting of coupled
ferromagnetic (FM) and antiferromagnetic (AF) layers is cooled in
field below the Neel temperature of its AF component, it shows the
asymmetric magnetization \cite{c,d,e,6,7,a}, which is referred to
as the exchange bias effect. Furthermore, the exchange bias field,
defined as the average of the two coercive fields, is observed to
decrease with increasing the number of the consecutive field
cycling, which is referred to as the training effect\cite{5}. The
exchange bias and training effects are very interesting and could
be used in future spintronics\cite{h,i,j} and data storage.
Usually, the FM layer is taken as a whole and the AF layer
consists of many grains. The AF grain is small enough to consists
of a single domain, and some uncompensated domains (or grains) may
be formed by defects or impurities\cite{15,k,16} and couple with
each other and with the FM domains. As the heterostructure is
cooled to a low temperature, the uncompensated spins in the grains
and domains become locked-in and prefer to a unidirection in the
interface, thus contribute to the magnetization shift\cite{7}.
Moreover, under the reversal of FM domains, the uncompensated
grains or domains will be irreversibly reorganized\cite{l,m,f,g,4}
and thus cause the training effect. The idea of domain states was
corroborated in some Monte Carlo simulations\cite{b}. On the other
hand, Hoffmann\cite{17} considered the biaxial anisotropy of the
AF sublattices and solved it by variational method. Actual
nonequilibrium dynamical properties of the magnetization are still
waiting to be elucidated. It is highly desirable and needed to
systematically investigate the two effects in a unified theory.

In this article we use a unified Monte Carlo dynamical
approach\cite{10} to study the FM/AF heterostructure in order to
investigate the exchange bias and training effect. Our simulated
result shows the obvious shift of hysteresis loops and the cycling
dependence of exchange bias. The magnetization of uncompensated AF
layer is still open after the field cycling is finished. The
exchange bias fields decrease with decreasing the cooling rate or
increasing the temperature and the number of the field cycling.
With the simulations, we shows the exchange bias can be
manipulated by controlling the cooling rate, the distributive
width of the anisotropy energy, or the magnetic coupling
constants. Essentially, these two effects can be explained on the
basis of the microscopically irreversible reversal of the AF
domains. More detailed results will be presented in the following.

The remaining part of this paper is organized as follows. In next
section we shall define our model and discuss our simulation
method. In section III we shall present our simulated results and
analysis. In section IV we shall discuss the microscopic mechanism
for the phenomena in a unified way. Finally, we shall give our
conclusion in section V.

\section{model and method}

According to experimental observations\cite{18}, for both
compensated and uncompensated AF layers the easy axis tends to
form along external cooling field direction rather than later
rotating field direction. In our model the AF layer consists of
many AF domains, and the FM layer consists of one single domain.
Assuming the cooling field is applied parallel to the AF/FM
interface, then all the easy axes of AF and FM domains lie in the
plane of the interface. The coupled bilayers of AF and FM domains
are shown in the inset of Fig. 1(a). The rectangles of the white
pattern represent the AF domains and the larger rectangle is the
single FM domain. The AF domains couple to each other
antiferromagnetically and the single FM domain couples to all the
AF domains ferromagnetically. We define the $z$ axis along the
common easy axis which lies in the interface plane. We apply the
external field to saturate the magnetization of the FM layer along
the $z$ axis.

For simplicity, we consider all the uncompensated spins in the AF
domains are the same. We use $S^\prime\vec{s}_i$ to denote the spin
vector of the $i$th AF domain and $S\vec{s}$ to denote that of the
single FM domain, where $S^\prime$ and $S$ are the uncompensated
spin values and FM spin respectively. Then we write the Hamiltonian
of the bilayers in an external field as
\begin{eqnarray}
H &=& -K_u (s^z)^{2} -\sum_i k_{ui} (s^z_i)^{2}
-\vec{B}\cdot(\gamma^\prime\sum_i\vec{s}_i +
\gamma \vec{s})\nonumber\\
& & +J_1\sum_{i,j}\vec{s}_i \cdot \vec{s}_j
-J_2\sum_i\vec{s}_i\cdot\vec{s}
\end{eqnarray}
where $\gamma^\prime=g \mu_0 \mu_BS^\prime$ and $\gamma=g
\mu_0\mu_B S$.  The first and second terms represents the
anisotropy of the FM domain and the AF ones, and $K_u$ and
$k_{ui}$ are the corresponding anisotropy constants. The third
term represents the Zeeman energy of the moments due to the
applied external field. The fourth term represents the
antiferromagnetic coupling among the AF domains. The last term
represents the ferromagnetic coupling between the FM and AF
domains.

Using $\theta_i$ and $\beta$ to describe the angles of the $i$-th
AF moment and the FM moment deviating from the common easy axis,
we can express the energies of the FM domain and the $i$-th AF as
\begin{equation}
 H^{\mathrm{FM}}=-(J_2\sum_is_is +K_u\cos\beta +\gamma B s)\cos\beta
\end{equation}
and
\begin{equation}
 H^{\mathrm{AF}}_i= (J_1s_i\sum_j s_j -J_2s_is {}
 -k_{ui}\cos\theta_i
 -\gamma^\prime B s_i)\cos\theta_i
\end{equation}
where both $s_i$ and $s$ are the scalars taking either 1 or -1.
Thus for the $i$-th AF domain the energy increment is $\Delta
E_i=k_{ui}\sin^2\theta_i-h_i(\cos\theta_i-1)$, where
$h_i=(-J_1\sum_js_j+J_2s+\gamma^\prime B)s_i$, and for the FM
domain the energy increment is $\Delta
E=K_u\sin^2\beta-h_F(\cos\beta-1)$, where
$h_F=(J_2\sum_is_i+\gamma B)s$. We can express $\Delta E$ and
$\Delta E_i$ as\cite{10}
\begin{equation}
 \Delta E=K_u[(1+\frac{h_F}{2K_u})^2-(\cos \beta+\frac{h_F}{2K_u})^2]
\end{equation}
and
\begin{equation}
 \Delta E_i=k_{ui}[(1+\frac{h_i}{2k_{ui}})^2-(\cos \theta_i+\frac{h_i}{2k_{ui}})^2]
\end{equation}
As a result, to reverse its moment, the the FM layer must
overcomes a barrier $E_b^F=K_u(1+h_F/2K_{u})^2$ if $|h_F|\leq
2K_{u}$, or $2h_F$ if $h_F> 2K_{u}$; and the $i$-th AF grain a
barrier $E_b^i=k_{ui}(1+h_i/2k_{ui})^2$ if $|h_i|\leq 2k_{ui}$, or
$2h_i$ if $h_i > 2k_{ui}$. If the condition $h_F < -2K_{u}$ or
$h_i < -2k_{ui}$ is satisfied, there is no barrier for the
reversal.

Actually, for the distribution of the AF anisotropy energy we use
a Gauss function, $f(k_{ui})=\exp[-(k_{ui}-k_u)^2/\sigma^2]$,
whose $\sigma$ and $k_u$ are set to 30.0 meV and 50.0 meV unless
stated otherwise. The anisotropy energy of the FM domain is set
200.0 meV without losing main physics. Thus the reversal rate for
a spin to reverse is $R=R_0e^{-E_b/k_BT}$, where $E_b$ is the
energy barrier and $R_0$ is the characteristic frequency. In our
simulations, $R_0$ is set to $1.0\times10^9$/s. We adopt a square
lattice for the AF domains and use $20\times20$ as its size. Since
we are only interested in the exchange bias and training effect at
the nanoscale, the AF lattice is enough to capture main physics.
Furthermore, we assume the AF domains have uniform moment 4.0
$\mu_B$ and the FM domain 2000 $\mu_B$. The coupling constant
$J_1$ is set to 4.0 meV, and $J_2$ 8.0 meV. In our simulations the
system is quenched from a high-enough temperature such as 610 K,
at which the AF layer is paramagnetic, to a low-enough temperature
such as 10 K. The magnetization and exchange bias fields are
calculated at the low temperature 10 K unless the temperature is
explicitly stated otherwise. The basic rate of changing
temperature is $\nu_0$= 50 K/s. The field sweeping rate is set to
0.5 T/s with the basic increment 0.1 T for each simulation step.

\section{Simulated Results and analysis}

At first, we let the AF/FM bilayers relax under a magnetic field
of 5.0 T at a high temperature 610 K. This temperature is enough
to make both the FM layer and the AF layer remain paramagnetic.
When the temperature decreases, the average magnetization values
of the two layers increases. The external field makes the average
magnetization of the FM layer have a large increase below 600 K,
and reach nearly to the saturated value at 500 K. When the
temperature becomes lower than 60 K, the average magnetization of
the AF layer looks like that of an antiferromagnet under an
applied field and is dependent on the cooling rate $\nu$. Then, we
further cool the bilayers under the same field. After the
temperature reaches down to 10 K, we start to change the field
while keeping the temperature unchanged. The field decreases from
5.0 T to -10.0 T and then increases back to 5.0 T for the first
hysteresis. Repeating the field cycling, we will make the second
hysteresis loop. The simulated results are shown in Fig. 1.
\begin{figure}[!htbp]
\includegraphics[width=8cm]{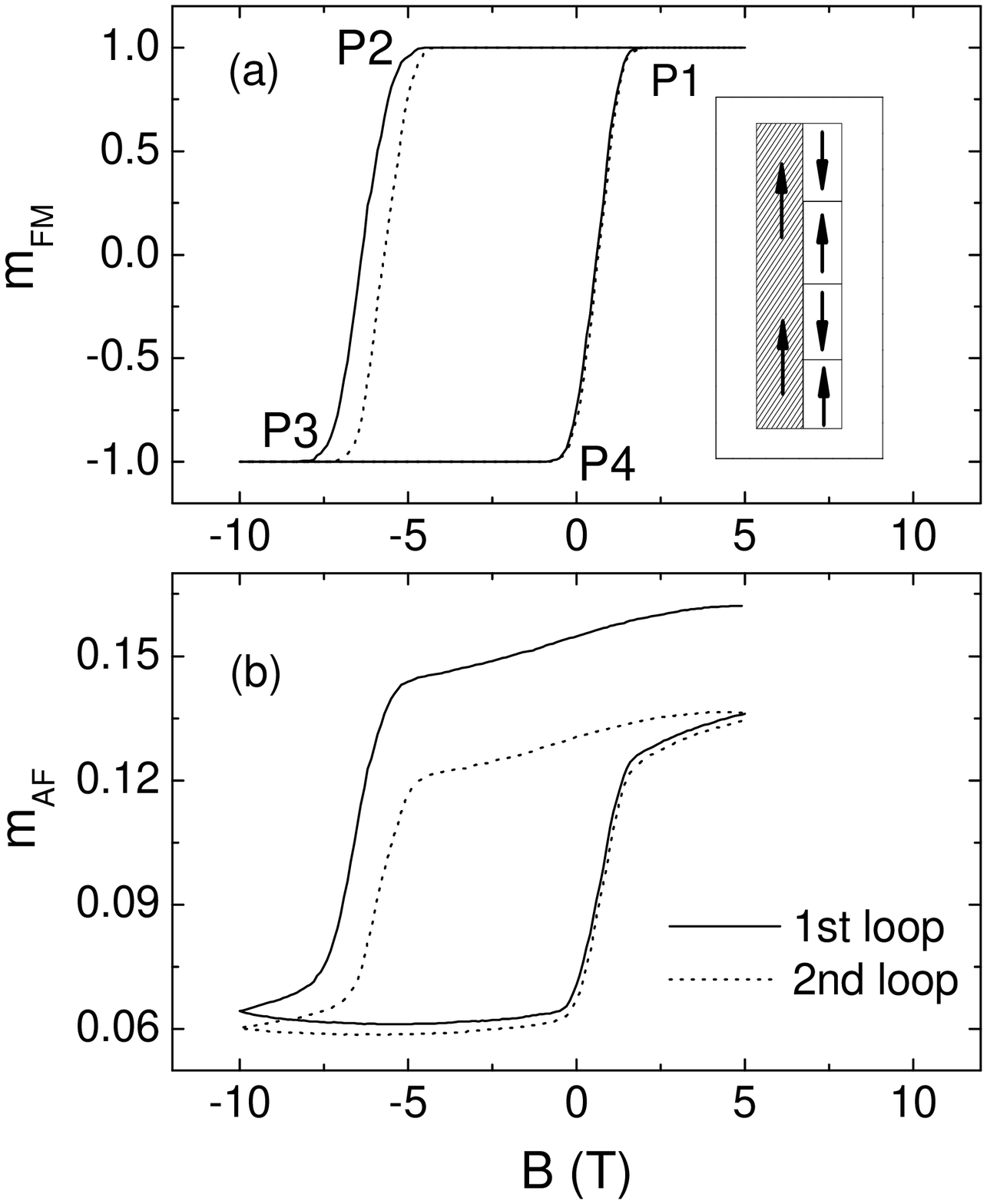}
\caption{\label{} The first two hysteresis loops of the FM (a) and
AF (b) layer at 10 K. The inset in (a) shows the AF/FM bilayers.
The hysteresis loop is obtained by changing the field in the order
of P1-P2-P3-P4-P1. }
\end{figure}

As shown in Fig. 1(a), the origin of the first hysteresis is
clearly shifted in the negative field direction and shows the
exchange bias. The exchange bias field is defined as
$H_E=(H_{cl}+H_{cr})/2$, where $H_{cl}$ and $H_{cr}$ is the
coercivity of the left and right branches. The left branch of the
second hysteresis moves towards the positive direction, but the
right branches of the first two loops almost coincide with each
other. Actually, any further loop almost does no difference in the
right branch from the second hysteresis. The shift of the second
loop clearly demonstrates that the bilayers magnetization depends
on the cycling history, which is known as training effect. Fig.
1(b) shows the magnetization of the AF layer, which drops largely
and opens widely due to the irreversible reversal of the AF
domains after the first field cycling is finished. The subsequent
magnetization is more smooth but still not closed, indicating the
continuing cycling dependence of exchange bias. This is consistent
with other Monte Carlo simulations\cite{b}.

\begin{figure}[!htbp]
\includegraphics[width=8cm]{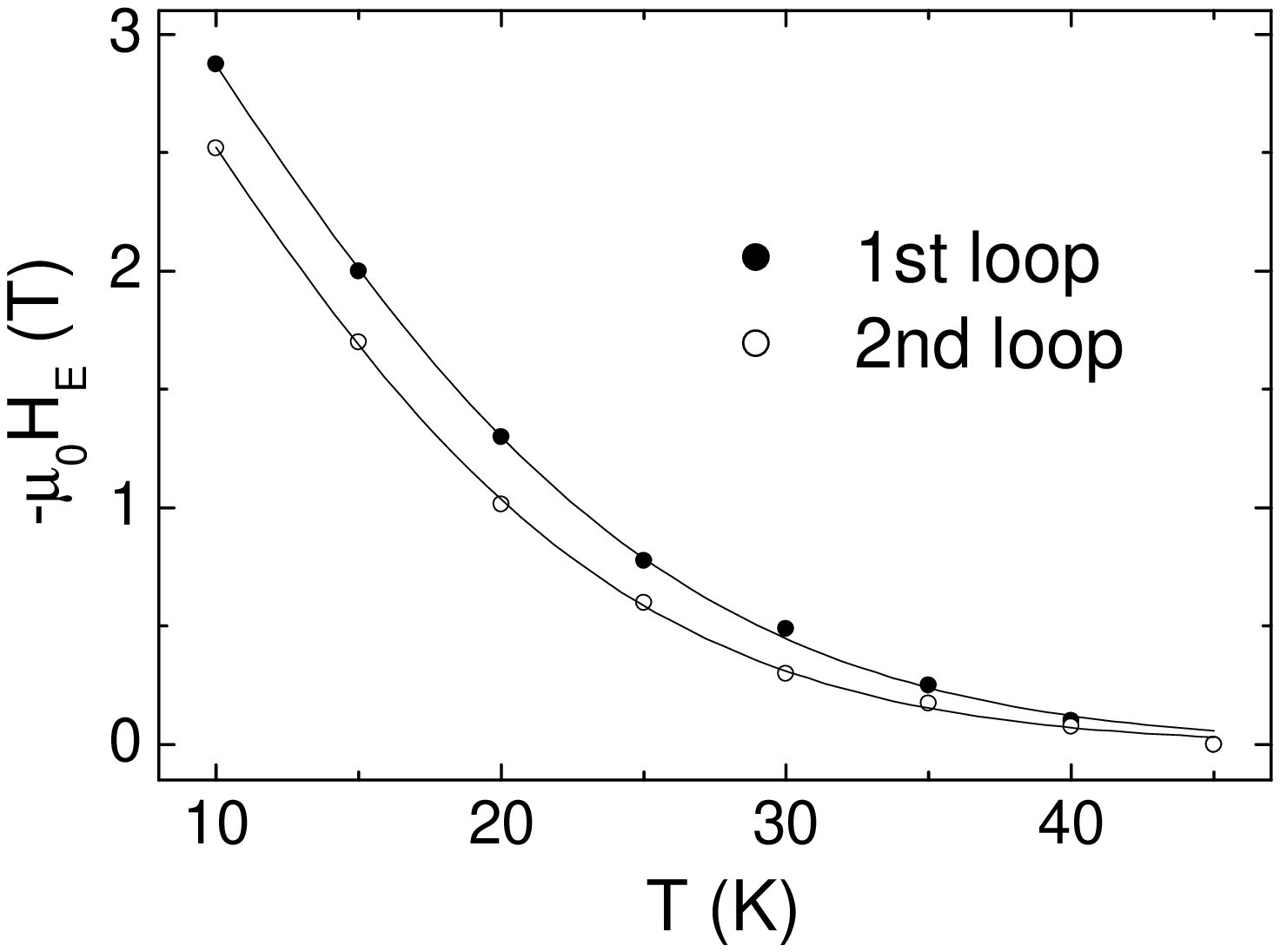}
\caption{\label{} Temperature dependence of the exchange bias
fields for the first two loops. The exchange bias field is
calculated at a given temperature after the system is cooled from
610 K to the temperature value. The lines are the fitting curves
in terms of the simple function defined in Eq. (6).}
\end{figure}

We study the effect of the temperature $T$ on the exchange bias
field, $H_E$, for different loops. Our simulated exchange bias
fields as functions of $T$ for the first two loops are shown in
Fig. 2. For both of the two curves, the data can be fitted by the
simple function
\begin{equation}
-\mu_0 H_E= a_1 e^{-(T/T_1)^{b_1}}
\end{equation}
where $a_1$, $T_1$, and $b_1$ are fitting parameters. For the
fitting in Fig. 2, the parameters $a_1$, $T_1$, and $b_1$ takes
4.26 T, 17.95 K, and 1.58 for the first loop, and 3.94 T, 16.66 K,
and 1.59 for the second loop. Our results are consistent with
experimental observation that the exchange bias field decreases
with increasing temperature \cite{19,k,n}.

\begin{figure}[!htbp]
\includegraphics[width=8cm]{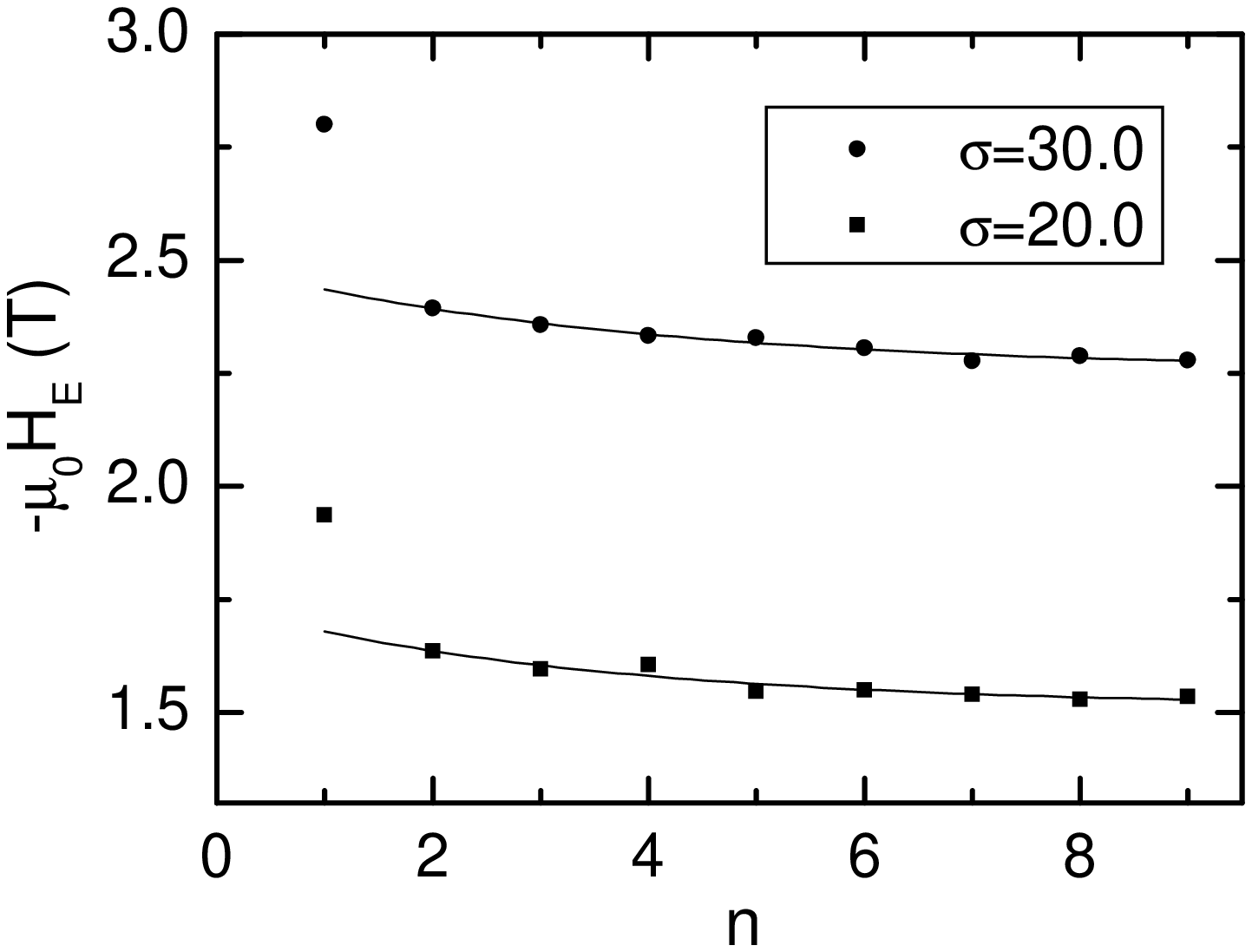}
\caption{\label{} The loop-number dependence of the exchange bias
fields for the Gaussian width $\sigma$=30.0 and 20.0 meV. The
temperature is 10 K. All the data except for $n=1$ can be well
fitted by a simple function $-\mu_0H_E(n)=a_2\rho^n+b_2$.}
\end{figure}

The exchange bias field is dependent on the field cycling number
$n$. Our simulated result from $n$=1 to $n$=9 is shown in Fig. 3.
Here, the temperature is kept at 10 K, and $\sigma$ is set to 20.0
and 30.0 meV. For both of the curves in Fig. 3, the data points
excepts of $n=1$ are well fitted by the simple function
$-\mu_0H_E(n)=a_2\rho^n+b_2$, where $a_2$, $b_2$, and $\rho$ are
the fitting parameters, taking 0.23 T, 2.26 T, and 0.76 for
$\sigma=30.0$ meV, and 0.22 T, 1.51 T, and 0.74 for $\sigma=20.0$
meV. The value of $-\mu_0H_E(1)$ usually is substantially above
the extrapolation of the other $-\mu_0H_E(n)$ ($n>1$). These
simulated results are in good agreement with experimental
observation\cite{5}.

\begin{figure}[!htbp]
\includegraphics[width=8cm]{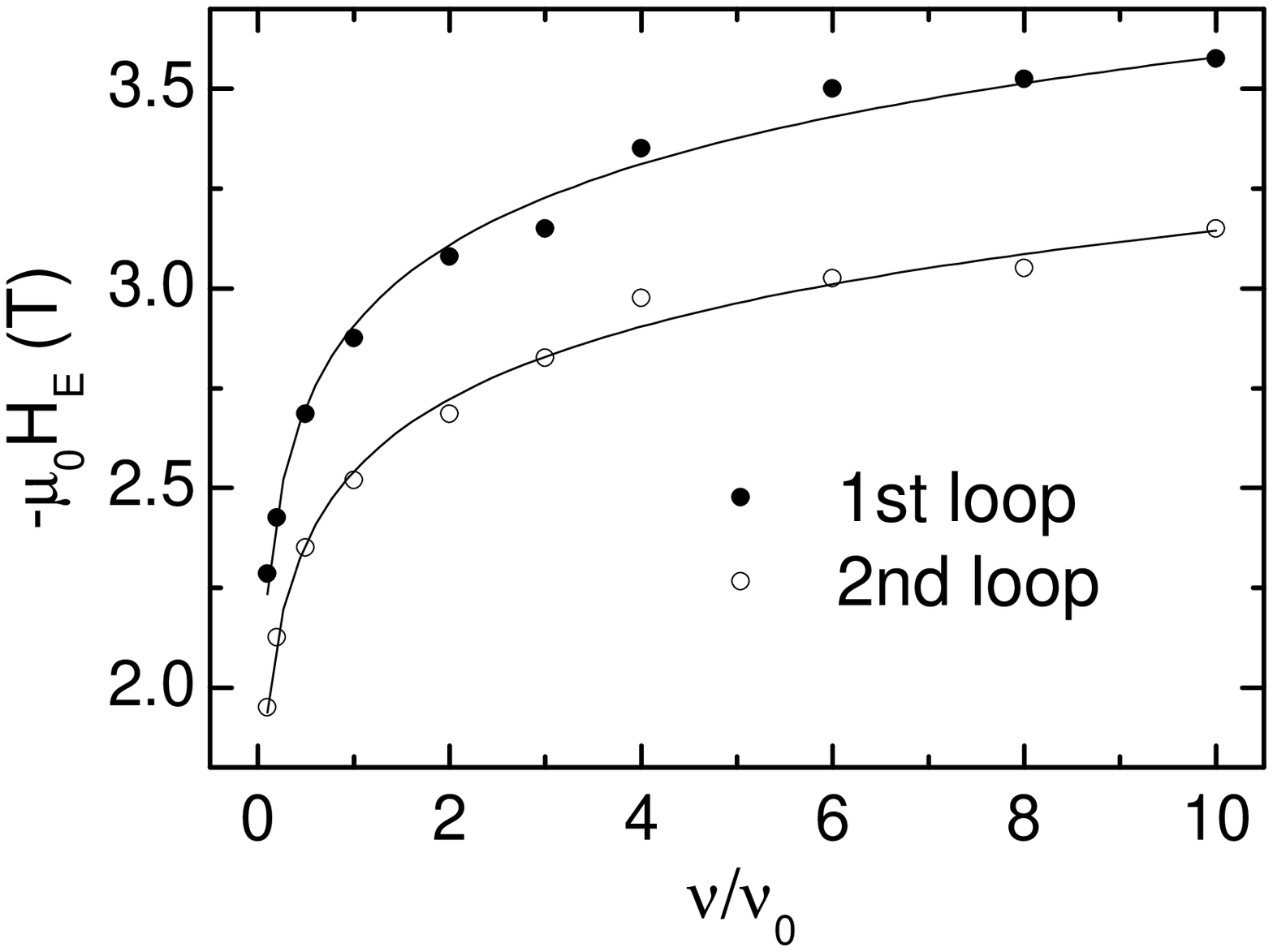}
\caption{\label{} The cooling-rate dependence of the exchange bias
fields for the first two loops. The lines are the fitting curves
in terms of Eq. (7). }
\end{figure}

Since the training effect reflects the non-equilibrium dynamical
magnetization which is caused by the irreversible reversal of
meta-stable domains formed during quenching, the quenching rate
must affect the exchange bias field. By changing the cooling rate
$\nu$, we study the exchange bias field as the function of
quenching rate $\nu$. The result is shown in Fig. 4. For both of
the loops, the data be well fitted by
\begin{equation}
  -\mu_0H_E=a_3\ln(b_3\frac{\nu}{\nu_0}+1)
\end{equation}
where $a_3$ and $b_3$ are 0.292 T and 20847 for the first loop,
and 0.262 T and 16248 for the second loop. The exchange bias field
at 10 K increases logarithmically with increasing the quenching
rate.

\begin{figure}[!htbp]
\includegraphics[width=8cm]{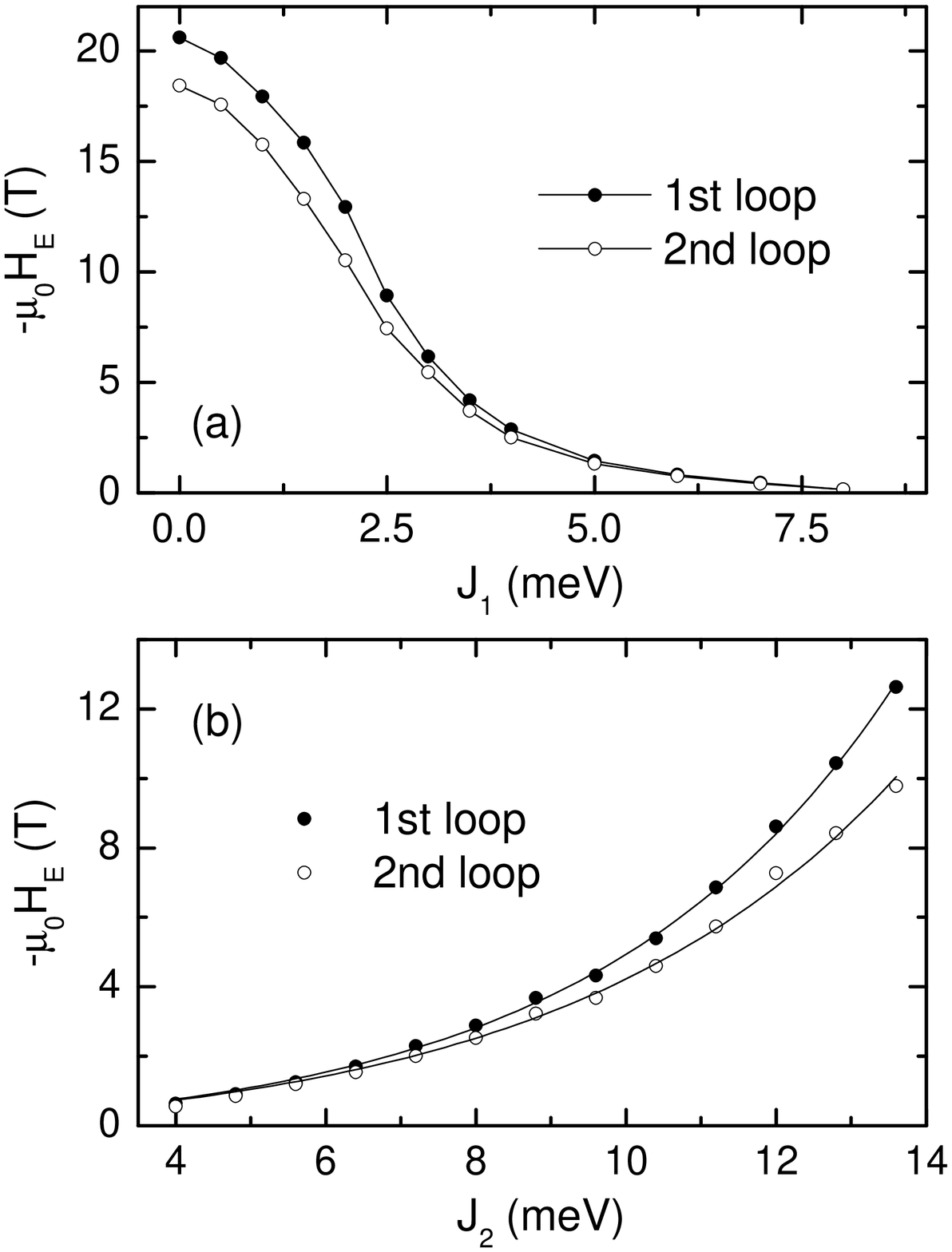}
\caption{\label{}The exchange bias fields as functions of the
coupling-constants $J_1$ (a) and $J_2$ (b) for the first two
loops. }
\end{figure}

\begin{figure}[!htbp]
\includegraphics[width=8cm]{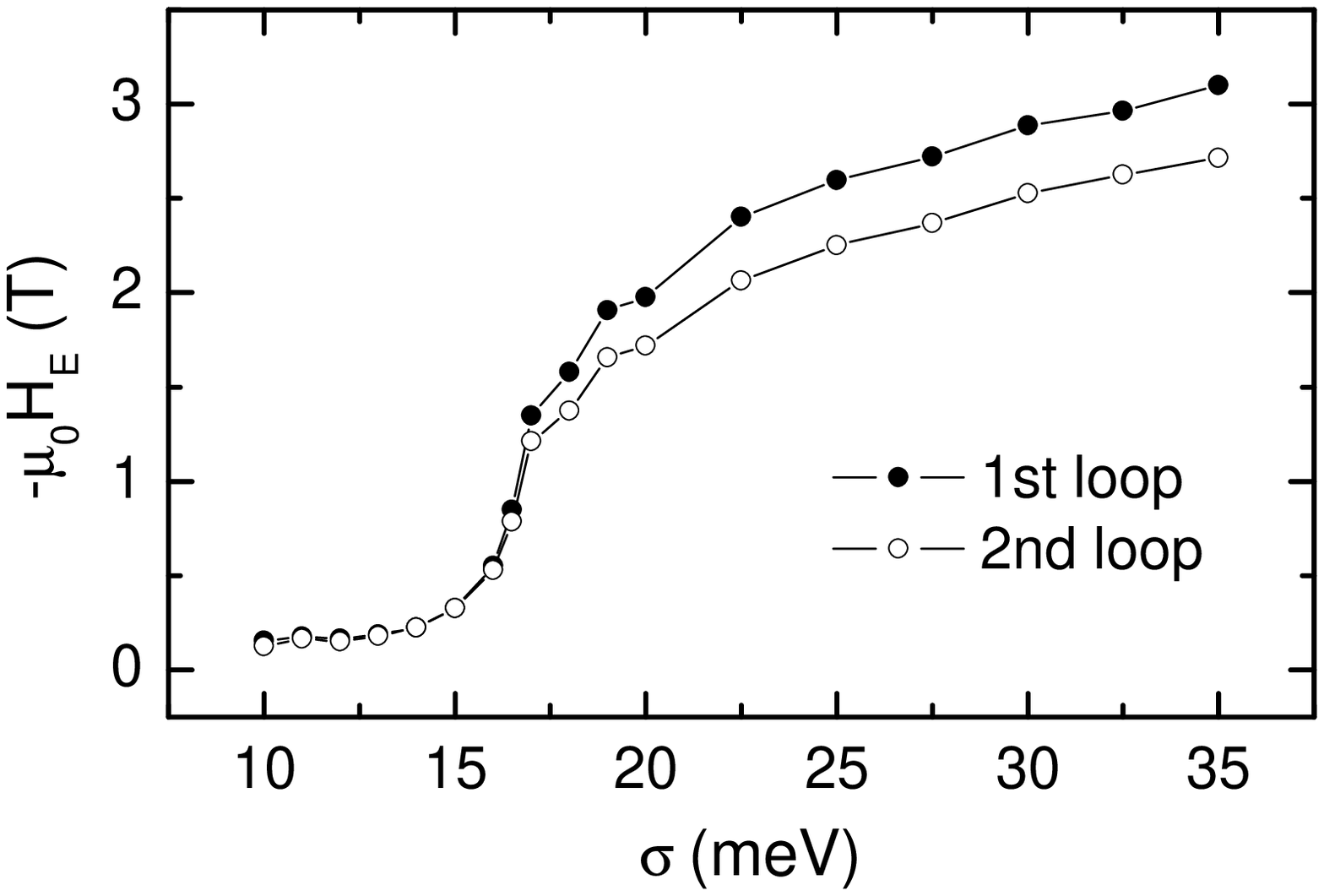}
\caption{\label{}The exchange bias field as a function of the
Gaussian width $\sigma$ for the first two loops. }
\end{figure}

It is interesting to investigate the dependence of the exchange
bias field on the coupling constants $J_1$ and $J_2$. The
simulated results are shown in Fig. 5. As shown in Fig. 5(a), the
exchange bias field decreases with increasing $J_1$. The training
effect is nearly unchanged when $J_1$ changes from 0 to 2meV, but
diminishes to zero quickly with increasing $J_1$ from 2meV. When
$J_1$ is larger than 6 meV, the exchange bias field already
becomes very small and the training effect is actually zero. In
contrast, the $J_2$ data points in Fig. 5(b) can be well fitted by
the simple function $-\mu_0H_E=a_4(\exp(J_2/b_4)-1)$, where $a_4$
and $b_4$ are the fitting parameters, taking 0.44 T and 4.01 meV
for the first loop, and 0.51 T and 4.49 meV for the second loop.
It is clearly shown that the exchange bias field increases
exponentially as $J_2$ increases. Both of the the exchange bias
field and the training effect can be enhanced by decreasing $J_1$
and increasing $J_2$, or by increasing $J_2/J_1$. This is
consistent with experimental trend\cite{p}. In Fig. 6 we shows how
the distributive width $\sigma$ of the AF anisotropy affects the
exchange bias fields. Clearly the exchange bias field increases
with $\sigma$, and so does the training effect. Experimentally,
the width can be increased by the additional nonmagnetic
impurities and the enhanced roughness of the AF crystalline
phases. This implies that the rougher the AF crystalline phases
are, the larger the exchange bias and training effect. Our result
reveals that the exchange bias field is determined by both the
coupling constants and the distributive width of the AF domain
anisotropy. These are useful to completely understand the
phenomenon\cite{p,o}.

\section{Trends and microscopic mechanism}

After being cooled down to the low temperature, the AF layer has a
non-zero net FM moment $M_A$ due to the driving of both the field
and the FM layer. Assuming there are $N_A$ AF domains, on average we
have the moments in part of all the $N_A$ AF domains aligning
parallel although they are coupled with AF interactions. The
exchange bias field is determined by the effective moment $M_A$, the
difference of $M_A$ between the first two loops determines the
training effect. Naturally, both the exchange bias field and the
training effect increase with increasing $J_2$ and with decreasing
$J_1$, as shown in Fig. 5. Actually, small $J_1$ does not affect the
effects, but larger $J_1$ than 2meV is harmful to the effects at 10
K. In addition, it is easily understood that $M_A$ decreases with
increasing the temperature $T$. As a result, both the exchange bias
field and the training effect decrease with increasing $T$, as shown
in Fig. 2. The exponential description in Eq. (6) reflects the fact
that moment reversals are thermally activated. It is reasonable that
both the exchange bias field and the training effect increase with
increasing the cooling rate $\nu$, as shown in Fig. 4. This is
mainly because the average magnetization of the AF layer increases
with $\nu$ when the temperature is below 50 K. When the cooling rate
approaches to zero, both the exchange bias field and the training
effect should be zero. In another word, our results should approach
to those of corresponding equilibrium systems when the cooling rate
$\nu$ approaches to zero.

As shown in Fig. 6, both the exchange bias field and the training
effect are nearly zero when the distributive width $\sigma$ of the
AF anisotropy energy is smaller than 15 meV, but they increase
substantially with increasing $\sigma$ for $\sigma > 15$ meV. This
means that the effects are dependent on a wide distribution of the
anisotropy energy. This can be understood in terms of the changing
of the energy barriers with the external field. From P1 to P2 in
Fig. 1(a), the effective barrier of the FM layer decreases but is
still high enough to avoid the reversal, but meanwhile, more and
more spins of the AF domains are reversed due to their lower
energy barriers. At the point P2, the FM moment is reversed with
the help of the field and the reversing of the AF domains. Anyway,
some AF domains with high energy barriers have their moments
unchanged, even after the FM layer has been reversed, and thus
there is a net average moment of the AF domains parallel to the
moment of the FM layer. This net average moment increases with the
distributive width $\sigma$. This explains the increasing of the
exchange bias field and training effect with increasing $\sigma$.
The more the field cycling loops, the longer the time. Actually,
this is similar to reducing the cooling rate $\nu$ in effect. As a
result, the exchange bias field decreases with increasing the
number of the field cycling. The turning point of the time scale
causes the largest drop happens between the first loops.

\section{Conclusion}

In summary, we use a unified Monte Carlo dynamical approach\cite{10}
to study the FM/AF heterostructure in order to investigate the
exchange bias and training effect. The magnetization of
uncompensated AF layer is still open after the first field cycling
is finished. Our simulated result shows the obvious shift of
hysteresis loops (exchange bias) and the cycling dependence of
exchange bias (training effect). The exchange bias fields decrease
with decreasing the cooling rate or increasing the temperature and
the number of the field cycling. With the simulations, we show the
exchange bias can be manipulated by controlling the cooling rate,
the distributive width of the anisotropy energy, or the magnetic
coupling constants. Essentially, these two effects can be explained
on the basis of the microscopical coexistence of both reversible and
irreversible moment reversals of the AF domains. Our simulated
results are useful to really understand the magnetization dynamics
of such magnetic heterostructures which should be important for
spintronic device and magnetic recording media \cite{p,o,20}. This
unified nonequilibrium dynamical method should be applicable to
other exchange bias systems.

\begin{acknowledgments}
This work is supported  by Nature Science Foundation of China
(Grant Nos. 10874232, 10774180, and 60621091), by the Chinese
Academy of Sciences (Grant No. KJCX2.YW.W09-5), and by Chinese
Department of Science and Technology (Grant No. 2005CB623602).
\end{acknowledgments}

\end{document}